\documentclass[iop]{emulateapj}
\usepackage{apjfonts}
\usepackage{graphicx}

\usepackage{color}
\usepackage{epstopdf}
\usepackage{amsmath}

\def\ltsima{$\; \buildrel < \over \sim \;$}
\def\simlt{\lower.5ex\hbox{\ltsima}}
\def\gtsima{$\; \buildrel > \over \sim \;$}
\def\simgt{\lower.5ex\hbox{\gtsima}}

\def\mb{m_\bullet}
\def\mc{m_c}

\def\tJ{t_\mathrm{J2000}}

\newcommand\lsim{\mathrel{\rlap{\lower4pt\hbox{\hskip1pt$\sim$}}
\raise1pt\hbox{$<$}}}
\newcommand\gsim{\mathrel{\rlap{\lower4pt\hbox{\hskip1pt$\sim$}}
\raise1pt\hbox{$>$}}}

\allowdisplaybreaks
\usepackage{color}

\shorttitle{Constraining Sgr A* companion}
\shortauthors{Naoz et al. }
\begin{document}

\title{  Constraining a companion of the galactic center black hole, Sgr A* } 

\author{   Clifford M. Will$^{1,2}$,   Smadar Naoz$^{3,4}$, Aur\'elien Hees$^5$,  Alexandria Tucker$^{1,6,7}$, Eric Zhang$^{8}$, Tuan Do$^3$, Andrea Ghez$^3$ }

\altaffiltext{1}{Department of Physics, University of Florida, Gainesville, FL 32611, USA}
\altaffiltext{2}{Institut d'Astrophysique, Sorbonne Universit\'e, 75014 Paris, France}
\altaffiltext{3}{Department of Physics and Astronomy, University of California, Los Angeles, CA 90095, USA}
\altaffiltext{4}{Mani L. Bhaumik Institute for Theoretical Physics, Department of Physics and Astronomy, UCLA, Los Angeles, CA 90095, USA}
\altaffiltext{5}{SYRTE, Observatoire de Paris, Universit\'e PSL, CNRS, Sorbonne Universit\'e, LNE, 61 avenue de l'Observatoire, F-75014 Paris, France}
\altaffiltext{6}{Department of Physics, University of Illinois, Urbana, IL 61801, USA}
\altaffiltext{7}{Illinois Center for Advanced Studies of the Universe,Department of Physics, University of Illinois, Urbana, IL 61801, USA}
\altaffiltext{8}{Department of Physics and Astronomy, University of California, Riverside, CA, 92507, USA}

\begin{abstract}
We use 23 years of astrometric and radial velocity data on the orbit of the star S0-2 to constrain a hypothetical intermediate-mass black hole orbiting the massive black hole Sgr A* at the Galactic center.  The data place upper limits on variations of the orientation of the stellar orbit at levels between $0.02$ and $0.07$ degrees per year.  We use a combination of analytic estimates and full numerical integrations of the orbit of S0-2 in the presence of a black-hole binary.  For a companion IMBH outside the orbit of S0-2 ($1020$ a.u.), we find that a companion black hole with mass $m_c$ between $10^3$ and $10^5 \, M_\odot$ is excluded, with a boundary behaving as $a_c \sim m_c^{1/3}$.  For a companion with $a_c < 1020$ a.u., a black hole with mass between $10^3$ and $10^5 \, M_\odot$ is excluded, with $a_c \sim m_c^{-1/2}$. These bounds arise from quadrupolar perturbations of the orbit of S0-2.  Significantly stronger bounds on an inner companion arise from the fact that the location of S0-2 is measured relative to the bright emission of Sgr A*, and that separation is perturbed by the ``wobble'' of Sgr A* about the center of mass between it and the companion.  The result is a set of bounds as small as $400 \, M_\odot$ at $200$ a.u.; the numerical simulations suggest a bound from these effects varying as $a_c \sim m_c^{-1}$.  We compare and contrast our results with those from a recent analysis by the GRAVITY collaboration.
\end{abstract}
\maketitle

\section{Introduction}
Sagittarius A* (Sgr A*) is a compact, bright radio source at the center of the Milky Way. Recent technological advances, such as the advent of adaptive optics (AO), have made it possible to observe stars orbiting this source.  The results imply that this is the likely location of a supermassive black hole (SMBH) of about 4 million solar masses \citep[e.g.,][]{Ghez+00,Ghez+08,Gillessen+09}, surrounded by a cluster of stars \citep[e.g.,][]{Ghez+03,Gillessen+09,Lu+13}. Combined infrared \citep[e.g., Keck observations,][]{Witzel+18}, radio and X-ray observations \citep[e.g., JVLA and Chandra observations,][]{Dibi+16,Capellupo+17} have revealed hot emission from gas near the event horizon of Sgr A*. 
Observations by the Event Horizon Telescope collaboration have provided evidence for the ``shadow'' of the black hole \cite[]{2022ApJ...930L..12E}. 
Thus, the proximity of the Milky Way's galactic center provides a unique laboratory for addressing issues in the fundamental physics of supermassive black holes, their impact on the central regions of galaxies, and their role in galaxy formation and evolution. 

The hierarchical nature of the galaxy formation paradigm suggests that galaxy mergers may result in the formation of {\em binaries} of SMBH \citep[e.g.,][]{DiMatteo+05,Hopkins+06,Robertson+06,Callegari+09}. While observations of SMBH binaries are challenging, there exist several confirmed binary candidates with sub-parsec to hundreds of parsec separations \citep[e.g.,][]{Sillanpaa+88,Rodriguez+06,Komossa+08,Bogdanovic+09,Boroson+09,Dotti+09,Batcheldor+10,Deane+14,Liu+14,Liu+17,Li+16,Bansal+17,Kharb+17,Runnoe+17,Pesce+18}.   
 Additionally,  observations of dual active galactic nuclei with kpc-scale separations have been suggested as SMBH binary candidates \citep[e.g.,][]{Komossa+03,Bianchi+08,Comerford+09bin,Liu+10kpc,Green+10,Smith+10,Comerford+18,Stemo+20}.

If Sgr A* is a member of a binary, could its companion be an intermediate-mass black hole (IMBH), that is, a black hole with a mass in the range of hundreds to thousands of solar masses?
 Recent observations by the LIGO/Virgo/Kagra collaboration have now confirmed the existence of 100 solar-mass black holes  \citep[e.g., GW190521][]{GW190521a+20,GW190521b+20}. Our galactic center may harbor IMBH as a result of a possible minor merger with a low-mass or dwarf galaxy or even with a globular cluster.  Such a scenario was considered by \citet{RM13}, who suggested that if IMBH serve as the seeds of SMBH in the center of galaxies, hierarchical galaxy evolution could yield many IMBH in our galaxy. Additionally, a combination of theoretical and observational arguments have led to speculation that IMBH may exist in the central parsec of the galaxy \citep[e.g.,][]{Hansen+03,Maillard+04,Grkan+20,Gualandris+09,Chen+13,Naoz+20,Generozov+20,Fragione+20,Zheng+20,Rose+22,Rose+23,Zhang+23}.

\begin{figure}
  \begin{center} 
    \includegraphics[width=\linewidth]{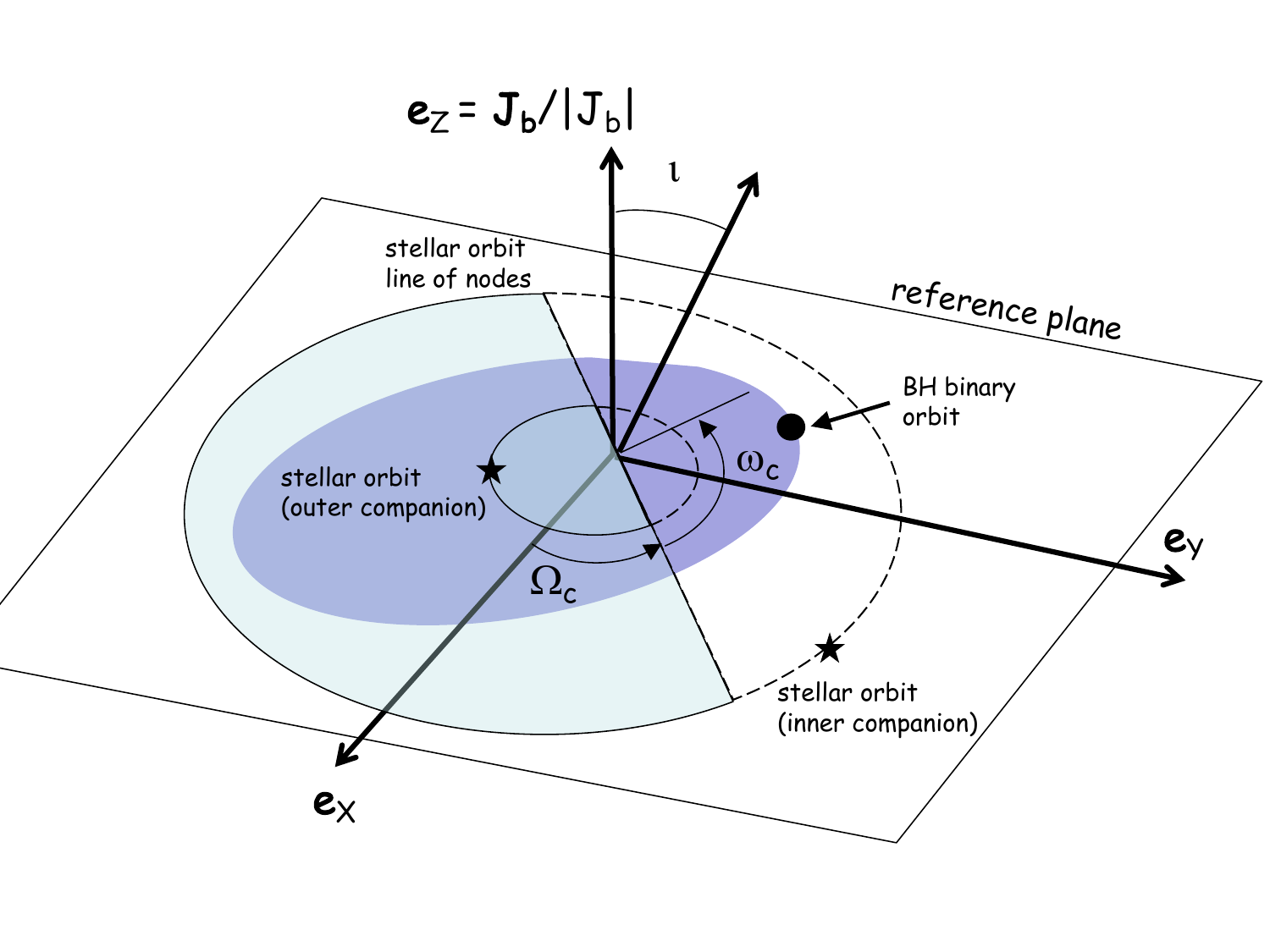}
  \end{center} 
  \caption{  \upshape   A hierarchical three-body system consisting of the Sgr A*-IMBH binary and the star S0-2.  The binary can be either inside or outside the orbit of the star   } \label{fig:Cartoon} 
\end{figure}

 In an earlier paper \citep{Naoz+20}, we constrained the allowable parameter space of an IMBH at the center of our Galaxy using the 20+ years of observations of the star S0-2, which orbits the SMBH Sgr A* with an orbital period of $16$ years and an eccentricity of about $0.88$.
 The recent closest approach of this star to Sgr A* (pericenter) has been  used to test and confirm the prediction of general relativity (GR) for the relativistic redshift  \citep[e.g.,][]{GRAVITY18,Do+19} and the advance of the pericenter 
\citep{Gravity+20}. The star S0-2 has been observed for more than two decades, and its orbit is sufficiently regular that, if there is a companion to  Sgr A*, it is either quite close to the main black hole, or well outside the orbit of S0-2, making this three-body system somewhat hierarchical in nature (Figure \ref{fig:Cartoon}). 

In this paper, we expand on \citet{Naoz+20} by developing improved analytical limits on a possible companion whose orbit is internal to that of  S0-2, and limits on a companion external to the orbit of S0-2, and by obtaining bounds on a companion using direct numerical fits using publicly available data on the orbit of S0-2 from the UCLA Galactic Center Group \citep{Do+19}.  In Sec.\ \ref{sec:analytical}, we use the equations of motion for hierarchical triple systems expanded to quadrupole order together with a suitable averaging procedure to produce analytic estimates of the bounds on an IMBH.  In Sec.\ \ref{sec:numerical}, we describe the numerical methods by which we obtain bounds using the full array of data on S0-2.  In Sec.\ \ref{sec:other}, we review and update other bounds on a hypothetical companion.  Section \ref{sec:conclusions} makes concluding remarks.

\section{Bounds on a companion IMBH: Analytic estimates  }
\label{sec:analytical}

We consider a hierarchical triple system consisting of the massive black hole Sgr A* and a lighter black-hole companion, with masses $m_{\bullet}$ and $m_c$, respectively, and a star of mass $m_\star$ such as S0-2
(see  Fig.\ \ref{fig:Cartoon}).  The two black holes may orbit each other within the orbit of the star, or the star may orbit the massive black hole in the presence of the lighter black hole orbiting outside the pair.  We will denote these cases as the inner and outer companion cases, respectively. 
In the {\it inner} companion case, we assume that the ratio $a_c/a$ of the inner and outer semimajor axes is small or that the inner orbital period is short compared to the stellar orbital period.  We treat the outer body, the star, as a massless test particle.  It has no effect on the inner binary, but its orbit is perturbed by the varying multipole moments of the inner binary's gravitational field.  These have been denoted ``inverse Eccentric Kozai-Lidov'' (iEKL) perturbations \citep[e.g.,][]{Naoz+17,Zanardi+17}.   For the case of an {\it outer} companion, the ratio $a/a_c$ and the ratio of the stellar orbit period to the period of the companion black hole are assumed to be small. The resulting dynamical evolution from this evolution is known as the ``Eccentric Kozai-Lidov'' \citep[EKL, e.g.,][]{Kozai,Lidov,Naoz16}.

We describe the three-body system in a coordinate system whose $Z$-axis is parallel to the system's total angular momentum (see Fig.\ \ref{fig:Cartoon}); since the star has negligible mass, this implies that the  $Z$-axis is perpendicular to the plane of the black-hole binary, which becomes the $X-Y$ reference plane (also called the ``invariable plane'').    The stellar orbit is inclined by an angle $\iota$ to the reference plane, intersecting it along the ``line of nodes'' at an angle $\Omega$ relative to the reference $X$ axis.  The equivalent line of nodes for the companion orbit makes an angle $\Omega_c = \Omega -  \pi $ relative to the $X$-axis.  The pericenter angle of the companion orbit is $\omega_c$ from the line of nodes, or $\Omega_c+\omega_c$ from the $X$-axis.  The star's pericenter is an angle $\omega$ from the ascending node of the stellar orbit.  Each orbit is characterized by semimajor axes $a_c$ and $a$ and eccentricities $e_c$ and $e$ as usual.
For simplicity, we assume that the two black holes have zero spin.  
 
The Newtonian equations of motion for the three-body system are given by
\begin{subequations}
\begin{align}
    \frac{d^2 \bf X}{dt^2} &= - \frac{G (\mb+m_\star)}{R^2}\mathbf N  - G\mc \left(\frac{\mathbf X - \mathbf x}{\left|\mathbf X - \mathbf x\right|^3}+\frac{\mathbf x}{r^3}\right) \, , 
     \label{eq:eomunexpanded}
   \\
    \frac{d^2 \mathbf x}{dt^2}&= - \frac{G (\mb +\mc)}{r^2} \mathbf n + G m_\star \left( \frac{\mathbf X - \mathbf x}{\left|\mathbf X - \mathbf x\right|^3} -\frac{\mathbf X}{R^3}\right) \, ,
    \label{eq:eombhbinary}
\end{align}
\label{eq:eomNewt}
\end{subequations}
where $\bf X$ and $\bf x$ are the positions of S0-2 and the companion, respectively, relative to the SMBH, with  $R=\left|\bf X\right|$, $r=\left|\bf x\right|$, $\mathbf{n} = \mathbf{x}/r$ and $\mathbf{N} = \mathbf{X}/R$.  For the purpose of our analytic estimates, we will henceforth set $m_\star = 0$; its mass of about $14 M_\odot$ will be included in the numerical integrations to be discussed in Sec.\ \ref{sec:numerical}.

However, here we must account for two observational subtleties. The first is that the observed astrometric position of S0-2 on the sky is defined using a reference frame attached to the bright emission at the location of Sgr A*; put differently, the observations are differential measurements between the two images.  The overall orientation of this reference frame is tied to  galactic masers \citep{sakai:2019ab}.   Now, in the absence of a companion black hole, this reference frame is the standard inertial frame attached to the center of mass of the Sgr A*-star system.  But if we are to consider a hypothetical companion, we must allow for the ensuing ``wobble'' of Sgr A* relative to the new center of mass located somewhere between the two black holes.   The most straightforward way to do this is to describe the orbit of S0-2 by a variable that aligns with the observations, namely
\begin{equation}
{\bf X} \equiv {\bf x}_\star - {\bf x}_\bullet \,.
\end{equation}
This is, in fact, the variable defined in Eq.\ (\ref{eq:eomunexpanded}).
The second is that the data on S0-2 also include radial-velocity (RV) measurements, which are defined relative to a local standard of rest, i.e., an inertial frame.  However, in the numerical analysis of the data, to be discussed in Sec.\ \ref{sec:numerical}, these measurements are consistently referred to the astrometric frame tied to Sgr A* by the transformation ${\bf V} \to  {\bf V} - (\mc/M) {\bf v}$,
where $M = \mb + \mc$.  Thus we will work in the astrometric reference frame throughout.

Depending on whether the star is inside or outside the orbit of the companion, we will expand Eq.\ (\ref{eq:eomunexpanded}) in powers of the small ratio of the two semimajor axes, treat the unperturbed orbital solution in terms of osculating orbit elements as described above, and insert the perturbing accelerations into the so-called Lagrange planetary equations which govern the variation of the orbit elements with time.

We wish to obtain the observable change of the orbit elements that characterize the {\em orientation} of the orbit of S0-2, namely $i$, $\Omega$ and $\omega$, that would be induced by the companion black hole.  Upper limits on these variations have been placed using data from the two observational groups;  the eccentricity $e$ of S0-2's orbit has also been well measured, leading to a rough upper bound on its rate of variation over the observation period, but we will not incorporate this in our study.

\subsection{Orbit elements in the reference frame and on the sky}

The star's angular orbit elements $\iota$, $\Omega$ and $\omega$ discussed
above  are defined in the reference system of Fig.\ \ref{fig:Cartoon}, and are {\em 
not} the same as the {\em observed} inclination, ascending node and pericenter 
angles ($i_{\rm sky}$ $\Omega_{\rm sky}$, $\omega_{\rm sky}$), defined with 
respect to the line of sight (${\bf e}_{Z,{\rm sky}}$) and a basis on the plane of 
the sky (${\bf e}_{X,{\rm sky}}$, ${\bf e}_{Y,{\rm sky}}$).   The transformation 
of the orbit elements between these two bases is quite complicated, particularly 
since we do not know the orientation of the orbital plane of the hypothetical 
companion black hole {\em a priori}.  However, there exist a set of invariant 
quantities that give a direct link between the two types of orbit elements.  These 
arise from $\bf {j}$ and $\bf {r}$, respectively the angular momentum and Runge-
Lenz unit vectors for the stellar orbit, defined in the system reference frame by 
\citep[see, e.g.,][]{PW2014}{}{}
\begin{align}
{\bf {j}} &\equiv \sin \iota (\sin \Omega \, {\bf e}_X - \cos \Omega \, {\bf e}_Y ) + \cos \iota \, {\bf e}_Z \,,
\nonumber \\
{\bf {r}} &\equiv (\cos \Omega \cos \omega - \cos \iota \sin \Omega \sin \omega ) \, {\bf e}_X
\nonumber \\
& \quad
+(\sin \Omega \cos \omega + \cos \iota \cos \Omega \sin \omega ) \, {\bf e}_X
+\sin \iota \sin \omega \, {\bf e}_Z \,.
\end{align} 
As these unit vectors are mutually orthogonal (${\bf j} \cdot {\bf r} = 0$), the resulting three degrees of freedom uniquely define 
$\iota$, $\Omega$ and $\omega$, and together with the unit vector ${\bf l} \equiv 
{\bf j} \times {\bf r}$ they define a basis of vectors for the stellar orbit that 
can be related by a suitable rotation to either the sky basis or the  reference 
basis defined by the companion black hole orbit.  However, the following scalar 
quantities, constructed from $d{\bf j}/dt$, $d{\bf r}/dt$, $d{\bf l}/dt$ and the 
three basis vectors are {\em invariant} under rotations (note that ${\bf j} \cdot 
d{\bf j}/dt  ={\bf r} \cdot  d{\bf r}/dt  ={\bf l} \cdot  d{\bf l}/dt  =0$):
 \begin{align}
{\cal I}_1 &=\frac{d{\bf r}}{dt} \cdot {\bf j} = -\frac{d{\bf j}}{dt} \cdot {\bf r}  = \sin \omega \frac{d\iota}{dt}  - \sin \iota \cos \omega \frac{d\Omega}{dt} \,,
\nonumber \\
{\cal I}_2 &= \frac{d{\bf j}}{dt} \cdot {\bf l} =-\frac{d{\bf l}}{dt} \cdot {\bf j}  = -  \cos \omega \frac{d\iota}{dt}  - \sin \iota \sin \omega \frac{d\Omega}{dt} \,,
\nonumber \\
{\cal I}_3 &=\frac{d{\bf r}}{dt} \cdot {\bf l} =-\frac{d{\bf l}}{dt} \cdot {\bf r}  = \frac{d\varpi}{dt} \,,
\label{eq:invariants}
\end{align}
where the variable $\varpi$ is defined by the relation $d\varpi/dt = d\omega /dt + \cos \iota \,  d\Omega /dt$.
Thus, a measurement on the sky of the variables that appear in Eqs.\ (\ref{eq:invariants}) yields, via construction of the three invariants, values of those invariants expressed in the reference system defined by the companion's orbit.

However, at present, apart from $d\omega/dt$ for the pericenter of S0-2 
\citep{Gravity+20}, which is dominated by the GR precession, no measurements of 
$d\Omega/dt$ or $d\iota/dt$ exist, only rms errors on their values, consistent 
with zero.   So in order to bound a hypothetical companion black hole, our 
strategy will be to convert the uncertainties in the rates of change $d\Omega/dt$ 
and $d\iota/dt$ into bounds, along with the uncertainty in $d\omega/dt$ after 
subtracting the GR effect.  For simplicity, we will use two invariants constructed 
from ${\cal I}_1$, ${\cal I}_2$ and ${\cal I}_3$, namely 
\begin{align}
\biggl |\frac{d{\bf j}}{dt} \biggr |^2 &={\cal I}_1^2 + {\cal I}_2^2   = \left ( \frac{d\iota}{dt} \right )^2 + \sin^2 \iota \left ( \frac{d\Omega}{dt} \right )^2 \,,
\nonumber \\
\biggl | \frac{d\varpi}{dt} \biggr |^2 &= {\cal I}_3^2  \,.
\end{align}
Because we have no {\em a priori} knowledge of the orientation of the companion 
black hole's orbit, it makes sense to marginalize or average these invariants over 
those orientations.  This implies averaging them over the two-sphere parametrized 
by $\iota$ and $\omega_c$  and the circle parametrized by $\omega$.  
 
 A similar approach was done in our previous work \citep{Naoz+20} for an inner 
 companion, averaging over the orbits.  As we highlight below, here we relax the 
 double averaging approach and expand this to the outer companion as well.  

\subsection{Outer companion}
\label{sec:outer}

We first consider a companion black hole outside the orbit of S0-2 so that S0-2 
and Sgr A* form the inner orbit of the hierarchical triple, and the companion is 
the outer perturber.    With $m_\star =0$ in Eq.\ (\ref{eq:eombhbinary}), $\bf x$  
evolves as a Keplerian orbit.  Expanding Eq.\ (\ref{eq:eomunexpanded}) in powers 
of $R/r$ through quadrupole order, we find the equation of motion for S0-2,
\begin{align}
\frac{d^2{\bf X}}{dt^2} &= - \frac{Gm_\bullet {\bf N}}{R^2}  + \frac{Gm_c R}{r^3} \left (3 ({\bf N}\cdot{\bf n}){\bf n}- {\bf N} \right ) 
 \,.
\label{eq:eomouterAst}
\end{align}

The first term in Eq.\ (\ref{eq:eomouterAst}) is the Keplerian acceleration of S0-
2 in the potential of the SMBH.  We define its osculating orbit elements in the 
usual manner, with ${\bf X} = R {\bf N}$, $R = a (1-e^2 )/(1+e \cos F)$, and with  
the unit vector ${\bf N}$ described in the $X-Y-Z$ basis using orbit elements 
$\iota$, $\Omega$ and $\omega$, along with sines and cosines of the true anomaly 
$F$, which satisfies the equation $dF/dt = \sqrt{G\mb a (1-e^2)}/R^2$ \citep[see][]
{PW2014}{}{}.   The second term is the conventional quadrupole perturbation due to 
the outer black hole.  
To obtain the evolution of the orbit elements, we drop the first, Keplerian term 
in Eq.\ (\ref{eq:eomouterAst}), and treat the remaining term as a perturbation of 
the Keplerian orbit; we find the components of the perturbations along the radial 
unit vector ${\bf N}$, perpendicular to that vector but in the orbital plane, and 
perpendicular to the orbital plane, and plug them into the Lagrange planetary 
equations.  

The conventional approach would be to carry out the double time average of the 
equations over the inner and outer orbits (also called the ``secular 
approximation'').  However, we are in a regime where the inner and outer orbital 
periods are not necessarily very different, making the secular approximation 
suspect.  In addition, we are less interested in detailed equations for the long-
term evolution of the orbit elements of S0-2 than in estimates for changes in its 
orientation over the orbit and a half corresponding to the actual observations.

Accordingly, our method will be as follows.  Holding the phase of the companion fixed
, we first integrate the planetary equations for S0-2's orbit from $-\pi$ to 
$2\pi$; this roughly corresponds to the $\sim24$ years of observation of S0-2 from 
apocenter through two pericenters. We divide by 1.5 orbital periods to get a per-
orbit rate of change.  We then construct the two invariants $|d{\bf j}/dt|^2$ and 
$|d\varpi/dt |^2$, and marginalize them over the unknown orbital phase of the 
companion and over the unknown inclination $\iota$ and the two unknown pericenter 
angles $\omega$ and $\omega_c$.  The results are
\begin{align}
\biggl | \frac{d{\bf j}}{d\tau} \biggr |^2 &= \zeta^2 {\cal A}_{\rm out} \alpha^{-6} \,,
\nonumber \\
\biggl |  \frac{d\varpi}{d\tau} \biggr |^2 &=  \zeta^2 {\cal B}_{\rm out} \alpha^{-6} \,,
\label{eq:secularouter}
\end{align}
where $\zeta \equiv m_c/\mb$, $\alpha \equiv a_c/a$, which in this case is greater than one;
$\tau$ is time measured in units of the stellar orbital period, $P = 2\pi (a^3/\mb)^{1/2}$; and 
\begin{align}
{\cal A}_{\rm out} &=\frac{(8+24e_c^2+3e_c^4)\left (81\pi^2(2+6e^2+17e^4)+512e^2(1-e^2) \right )}{1080(1-e_c^2)^{9/2} (1-e^2)} \,,
\nonumber \\
{\cal B}_{\rm out} &=\frac{(8+24e_c^2+3e_c^4)\left (64(1-2e^2)^2+567\pi^2 e^2(1-e^2) \right )}{270(1-e_c^2)^{9/2} e^2} \,,
\end{align}
where $e_c$ is the eccentricity of the companion black hole's orbit.

\subsection{Inner companion}
\label{sec:inner}

We next consider a companion whose orbit is inside that of S0-2, so that $a_c < a$.   Expanding Eq.\ (\ref{eq:eomunexpanded}) (with $m_\star = 0$)  in powers  of $r/R$ we obtain the equation of motion 
\begin{align}
\frac{d^2{\bf X}}{dt^2} &= - \frac{GM{\bf N}}{R^2} -  \frac{Gm_c{\bf n}}{r^2} -\frac{Gm_c r}{R^3} \left (3 ({\bf N}\cdot{\bf n}){\bf N}- {\bf n} \right ) 
\nonumber \\
& \quad -  \frac{3Gm_c r^2}{2R^4} \left (5 ({\bf N}\cdot{\bf n})^2 {\bf N} - 2({\bf N}\cdot{\bf n}){\bf n} - {\bf N} \right ) \,.
\label{eq:eominnerAst}
\end{align}
The first term in Eq.\ (\ref{eq:eominnerAst}) is the Keplerian acceleration of the star in the asymptotic field of the binary of mass $M$.  
 The second term is the ``fictitious'' acceleration of S0-2 caused by the 
 acceleration of the astrometric reference frame attached to Sgr A* which 
 ``wobbles'' around the center of mass of the inner binary, while the third term 
 is a ``dipole'' term caused by the fact that the vector $\bf X$ no longer points 
 toward the center of mass of the black hole binary.  The final term is the 
 conventional quadrupolar perturbation, except for the fact that the mass factor 
 in Eq.\ (\ref{eq:eominnerAst}) is $m_c$ instead of the usual reduced-mass factor 
 $\mu = m_c m_\bullet /M$.

Carrying out the same procedure as for the outer companion, we obtain   
\begin{align}
\biggl |\frac{d{\bf j}}{d\tau}\biggr |^2 &= \frac{\zeta^2}{(1+ \zeta)^2}  \left ( {\cal A}_{\rm in} \alpha^4+ {\cal B}_{\rm in} \alpha^2 + {\cal C}_{\rm in} \alpha^{-1} + {\cal D}_{\rm in} \alpha^{-4} \right ) \,,
\nonumber \\
\biggl |\frac{d\varpi}{d\tau}\biggr |^2 &= \frac{\zeta^2}{(1+ \zeta)^2} \left ( {\cal E}_{\rm in} \alpha^4+ {\cal B}_{\rm in} \alpha^2 - {\cal C}_{\rm in} \alpha^{-1} + {\cal F}_{\rm in} \alpha^{-4} \right ) \,,
\label{eq:secularinner}
\end{align}
where $\alpha \equiv a_c/a$, and 
\begin{align}
{\cal A}_{\rm in} &\equiv \frac{(8+40e_c^2+15e_c^4)(81\pi^2+16 e^2)}{540 (1-e^2)^4} \,,
\nonumber \\
{\cal B}_{\rm in} &\equiv \frac{8(2+3e_c^2)}{27(1-e^2)^2} \,,
\nonumber \\
{\cal C}_{\rm in} &\equiv \frac{32}{27(1-e^2)} \,,
\nonumber \\
{\cal D}_{\rm in} &\equiv \frac{(2+e_c^2)\left [ 16(1-e^2)+81\pi^2 e^2 \right ]}{54(1-e_c^2)^{5/2} (1-e^2)} \,,
\nonumber \\
{\cal E}_{\rm in} &\equiv \frac{(8+40e_c^2+15e_c^4) \left [16(1+2e^2)^2+243 \pi^2 e^2 \right ]}{540(1-e^2)^4} \,,
\nonumber \\
{\cal F}_{\rm in} &\equiv \frac{(2+e_c^2)\left [ 81\pi^2 (1-e^2)+ 16e^2 \right ]}{54 (1-e_c^2)^{5/2} e^2} \,.
\end{align}

The effects of the terms in the equation of motion (\ref{eq:eominnerAst}) can be seen in these expressions, with the quadrupole term ($\sim r^2$)  appearing (squared) in the $\alpha^4$ term, the dipole term ($\sim r$) appearing in the $\alpha^2$ term, the wobble effect ($ \sim r^{-2}$) appearing in the $\alpha^{-4}$ term, and the $\alpha^{-1}$ term representing a cross-term between the wobble and dipole effects.

However, the step where we integrated over the orbit of S0-2 while holding the companion fixed in its orbit, while perhaps not unreasonable when the orbital periods are comparable ($a_c \sim a$), becomes problematic when the period of the companion is much shorter than that of the star.  In this situation, the wobble and dipole effects on the orbit elements actually average to zero over one orbit of the companion.  We will discuss this in detail when we compare these estimates with the data in the next subsection.  

\begin{figure}[t]
  \begin{center}
    \includegraphics[width=\linewidth]{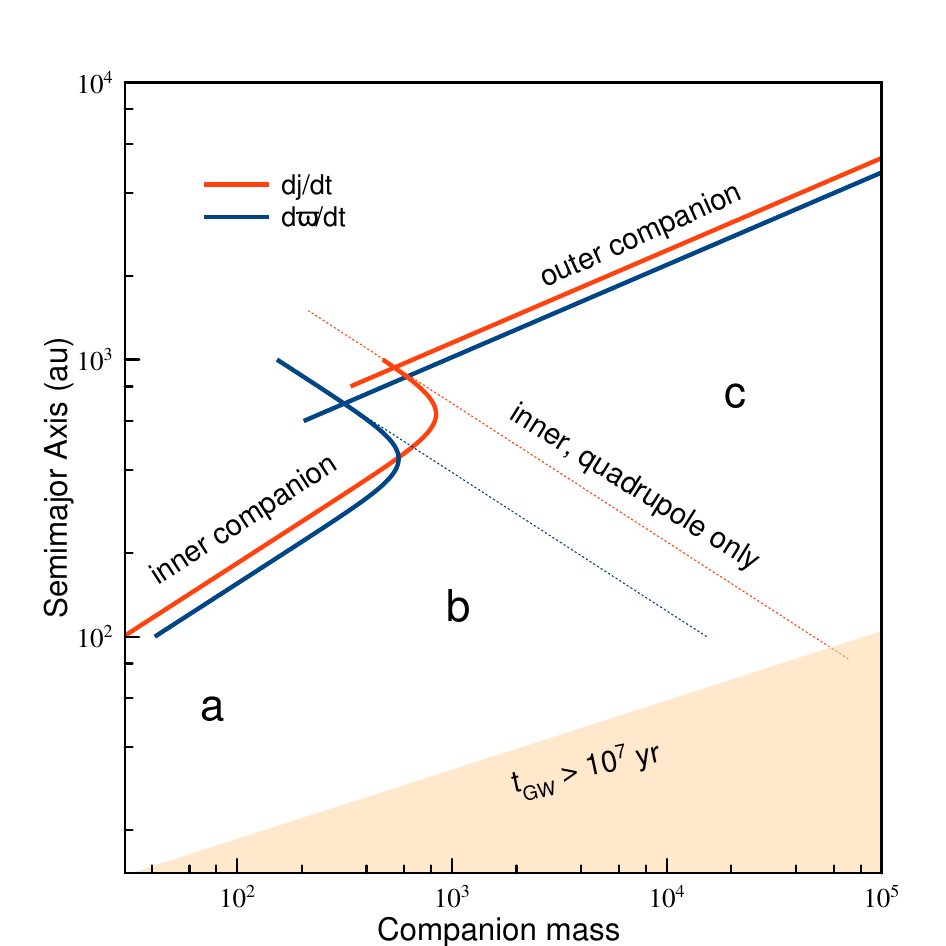}
  \end{center} 
  \caption{  \upshape Bounds on the mass and semimajor axis of a companion IMBH in an orbit with $e_c = 0.6$.  Red: bounds from $|d{\bf j}/dt|$; Blue: bounds from $|d\varpi/dt|$.  Dotted lines denote bounds on an inner companion purely from quadrupole perturbations of  $|d{\bf j}/dt|$ and $|d\varpi/dt|$.  The regions labelled ``a'', ``b'' and ``c'' are discussed in Sec.\ \ref{sec:constraints}.  The tan area denotes a gravitational-wave damping timescale for the companion shorter than $10^7$ years.  } \label{fig:Bounds} \vspace{0.2cm}
\end{figure}

\subsection{Observational constraints}
\label{sec:constraints}

With over 23 years of astrometric measurements and 19 years of radial velocity measurements, the Galactic Center Group has measured the orbit elements of S0-2 with reasonable accuracy \citep{Do+19} (see Table \ref{tab:obsbounds}).  The GRAVITY collaboration has measured the orbit elements of the star (denoted by them as S-2) with similar uncertainties \citep{GRAVITY18}.  Apart from a published measurement of $d\omega/dt$, which agrees with the GR prediction \citep{Gravity+20}, no significant change in the other orbit elements has been detected to date.   
 
In addition, the publicly available data from \cite{Do+19} has made it possible to estimate upper limits on a linear drift for each of S0-2's orbital elements. The orbital fit methodology is described in the Supplementary Materials of \cite{Do+19}. The parameters included in the orbital fit were the mass of Sgr A*, its distance and line-of-sight velocity and its position and velocity on the plane of the sky, the 6 standard orbital elements for S0-2. In addition, a linear drift for each orbital parameter was included.  Statistical tests for model selection based on Bayesian evidence \citep[see][]{Do+19} showed that no significant deviations from zero were measured. An estimate of the 95 \% upper limit on a linear drift of S0-2's orbital elements was derived from the posterior probability distribution of the fit combined with an estimate of the systematic uncertainty derived from a ``jackknife'' analysis at the level of the reference frame construction \citep[see][]{boehle:2016wu,sakai:2019ab,Do+19}. As a result, a 95\% confidence upper limit on $|d \omega_{\rm sky} /dt|$ was reported in \cite{hees:2017aa}, and bounds on $| d\Omega_{\rm sky} /dt |$ and   $| di_{\rm sky} /dt |$ were reported in \citet{Naoz+20}.  These are summarized in Table \ref{tab:obsbounds}.  The analyses described above did not include the possibility of an IMBH companion; that will be the subject of the next section.

\begin{table}[t]
\begin{center}
\caption{Orbit elements on the sky of Sgr A* (from Table 1 of \cite{Do+19}) and bounds on their variations (from \cite{hees:2017aa} and \cite{Naoz+20}). The quantities $\sigma_{\rm Stat}$ and $\sigma_{\rm Syst}$ denote the statistical and systematic errors on the elements } 
\begin{tabular}{l@{\hskip 0.5cm}c@{\hskip 0.5cm}c@{\hskip 0.5cm}c@{\hskip 0.5cm}c@{\hskip 0.5cm}}
\hline
Orbit&Value&$\sigma_{\rm Stat}$&$\sigma_{\rm Syst}$&Bound on \\
 element&&&&variation\\
\hline 
$e$&$0.8858$&$0.0004$&$2.8 \times 10^{-5}$&$2.9 \times 10^{-4}$ yr$^{-1}$\\
$\iota$ (deg)&$133.82$&$0.18$&$0.13$&$0.02$ deg/yr\\
$\omega$ (deg)&$66.11$&$0.24$&$0.077$&$0.07$ deg/yr\\
$\Omega$ (deg)&$227.49$&$0.29$&$0.11$&$0.07$ deg/yr\\
\hline
\end{tabular}
\label{tab:obsbounds}
\end{center}
\end{table}

Using the observed sky-basis orbit elements for S0-2, $a = 1020$ au, $e = 0.886$, $\iota = 134^{\rm o}$, and the estimates shown in Table \ref{tab:obsbounds}, we construct the observed bounds on the invariants $|d{\bf j}/dt |$ and $|d\varpi/dt|$.  Combining these with Eqs.\ (\ref{eq:secularouter}) and (\ref{eq:secularinner}), we obtain the bounds plotted in Fig.\ \ref{fig:Bounds}.  We have not utilized the bound on variations in the eccentricity in part because, for an inner companion, the eccentricity is constant to quadrupole order, while for an outer companion, the bound  is not significantly different from the bounds obtained using the angular invariants, so we have not displayed that bound in Fig.\ \ref{fig:Bounds}.  The tan area in Fig.\ \ref{fig:Bounds} denotes the region where gravitational-wave emission would have caused a companion to merge with Sgr A* within the $\sim 10$ million-year age of S0-2.

The red and blue lines in Fig.\ \ref{fig:Bounds} labeled ``outer companion'' show the $a_c \sim m_c^{1/3}$ trend characteristic of quadrupolar perturbations of the orbit of S0-2.  Companions below and to the right of those curves are excluded.  The lines labeled ``inner companion'' begin with the $a_c \sim m_c^{-1/2}$ dependence expected for quadrupole perturbations and are extended as dotted lines labeled ``inner, quadrupole only''.  The bounds indicated by the dotted lines are largely consistent with the bounds shown in our earlier paper \citep{Naoz+20}.  The region labeled ``c'' between the solid and dotted lines is excluded because of quadrupolar perturbations. However, as $a_c$ decreases, the effect of the wobble of Sgr A* begins to dominate, and the curves bend over to display an $a_c \sim m_c^{1/2}$ dependence.  One might be tempted to conclude that a large set of companions below the lines labeled ``inner companion'' are excluded.  However, we must recall that these curves were obtained by holding the companion fixed in its orbit while integrating the perturbations over the 1.5 orbits of S0-2.  In fact, the companion's period is shorter than that of the star, and therefore we should have integrated over the companion's orbit first, as called for in the secular approximation.  In that case, the wobble and dipole contributions from Eq.\ (\ref{eq:eominnerAst}) integrate precisely to zero.   In reality, over the 1.5 orbits of S0-2, the perturbations due to the wobble and dipole terms will not average precisely to zero, but will be suppressed relative to what is implied by Eqs.\ (\ref{eq:secularinner}), depending on the specific relation between the orbital periods.  In other words, the bound labeled ``inner companion'' could be porous, such that, for smaller $a_c$, some companions might be allowed because their wobble and dipole perturbations are sufficiently suppressed.  This corresponds to the region labeled ``a''.  On the other hand, for a given $a_c$, the wobble and dipole effects grow linearly with $m_c$, and thus, for  higher-mass companions, the wobble and dipole perturbations will be larger and more likely to lead to exclusion, corresponding to the region labeled ``b''.

This long discussion illustrates the difficulty of drawing firm analytic conclusions in regimes where the hierarchical assumption is only marginally true, particularly when the frequency of the perturbation is shorter than that of the orbit being investigated.  In such a case, one must turn to full numerical integrations of the equations of motion in hopes of obtaining a truer picture.   Those will be the subject of the next section.

\section{Bounds on a companion IMBH: Inference using Galactic Center data and integration of the 3-body equations of motion}
\label{sec:numerical}

In this section we carry out a full numerical integration of the equations of motion of S0-2 in the presence of a hypothetical IMBH companion and perform a Bayesian exploration of the parameter space using the publicly available data from UCLA's Galactic Center Group. The model fitted to the data is new and will be described in detail in this section while the methodology and data used in our analysis are identical to the ones from \cite{Do+19} and will be briefly summarized.

In this analysis, we consider the Newtonian motion of the three bodies (the SMBH, its possible companion and the star S0-2) but we also include $1/c$ relativistic  correction in the expression of the radial velocities (relativistic redshift). In total, the model depends on 20 free parameters:
\begin{itemize}
    \item the masses of the SMBH $\mb$ and of its companion $\mc$. 
    \item six parameters describing the initial conditions of the companion with respect to the SMBH. We use osculating elements at the reference epoch J2000: semi-major axis $a_c$, eccentricity $e_c$, inclination $i_c$, argument of periastron $\omega_c$, longitude of ascending node $\Omega_c$ and mean anomaly at J2000 $m_{0,c}$. 
    \item six parameters corresponding to the initial conditions of S0-2. We use osculating elements at the reference epoch J2000: orbital period $P$, eccentricity $e$, inclination $i$, argument of periastron $\omega$, longitude of ascending node $\Omega$ and time of closest approach $t_0$.
    \item the distance $R_0$ between the Solar System and the Galactic Center.
    \item four parameters to parametrize a possible drift of the reference frame ($x_0$, $y_0$, $v_{x_0}$, $v_{y_0}$, see \cite{Do+19}).
    \item one offset $v_{z_0}$ for the radial velocities (RV).
\end{itemize}

In the code, we fix the mass of the star S0-2 $m_\star$ to 13.6$M_\odot$, the nominal value estimated by \citet{habibi:2017aa}. The osculating elements from both the BH companion and S0-2 are transformed into cartesian positions and velocities at J2000 using regular Keplerian transformations. Note that for S0-2, we distinguish two cases: (i) if $a_\mathrm{c} >$ 1000 a.u., we use the mass of the SMBH to convert the osculating elements into cartesian coordinates since in that case, S0-2 is orbiting around the SMBH and it is perturbed by the outer body and (ii) if $a_\mathrm{c} <$ 1000 a.u., the transformation between osculating elements and Cartesian coordinates is performed using the total mass of the binary system since in this case, S0-2 is orbiting around the center of mass of the binary system (see Eqs.\ (\ref{eq:eomouterAst}) and (\ref{eq:eominnerAst})).

From the cartesian coordinates and velocities at J2000, we integrate the Newtonian equations (\ref{eq:eomNewt}) for the three body system.  We safely neglect the first post-Newtonian corrections to these equations of motion, considering that the dataset used in this analysis is not sensitive to these; see the discussion in \citet{Do+19}. We integrated these equations of motion from J2000 forward and backward in time in order to cover the full observational time span, i.e., from 1995 to 2018.

From the results of the numerical integration, we compute both the astrometric and the RV observable. We take into account the R\"omer time delay, which is due to the fact that the speed of light is finite, and thus the signal from the star takes a certain amount of time to propagate through S0-2's orbit in the $Z$-direction. To first order in $1/c$, this delay can be approximated by \citep{Do+19}
\begin{equation}
    t_\mathrm{em} = t_\mathrm{obs} - \frac{Z(t_\mathrm{obs})}{c} \, ,
\end{equation}
where $t_\mathrm{obs}$ is the epoch of observation, $t_\mathrm{em}$ the epoch of emission of the light and $Z(t)$ is the third component of $\mathbf X(t) = \left(X,Y,Z\right)$.

The astrometric observations are the relative sky position of S0-2 with respect to the SMBH, i.e.,
\begin{subequations}
    \begin{align}
        \bar X(t_\mathrm{obs}) &= \frac{X(t_\mathrm{em})}{R_0}+x_0 +v_{x_0}\left(t_\mathrm{obs}-\tJ\right) \, , \\ 
        \bar Y(t_\mathrm{obs}) &= \frac{Y(t_\mathrm{em})}{R_0}+y_0 +v_{y_0}\left(t_\mathrm{obs}-\tJ\right)\, ,
    \end{align}
\end{subequations}
where $(\bar X,\bar Y)$ are the astrometric observables and $x_0, y_0, v_{x_0}$ and $v_{y_0}$ model a 2D offset and linear drift of the reference frame.

On the other hand, the RV is not defined with respect to the SMBH but rather to the center of mass of the system. This is due to the fact that the RV is defined with respect to the local standard of rest \citep{1998MNRAS.298..387D,2016ARA&A..54..529B}. More precisely, the RV observable is computed as 
\begin{equation}
    RV\left(t_\mathrm{obs}\right) = V_z\left(t_\mathrm{em}\right) - \frac{\mc}{\mb+\mc} v_z\left(t_\mathrm{em}\right) + v_{z_0} + v_{GR}\, , 
\end{equation}
where $V_z = {dZ}/{dt}$ (the line-of-sight component of the velocity of S0-2 with respect to the SMBH), $v_z={dz}/{dt}$ (the line-of-sight component of the velocity of the companion with respect to the SMBH) and $v_{z_0}$ is a constant velocity offset that accounts for possible systematic effects in the radial velocity measurement or in the VLSR correction. The first term in this equation is the standard Newtonian velocity projected along the line-of-sight. The second term is a correction to take into account the fact that the measured RVs are expressed in the local standard of rest (i.e., the origin of the RV reference frame is the dynamical center of mass of the binary black-hole system and this term corrects for the motion of the SMBH with respect to the center of mass of the system). This contribution becomes non-negligible only for heavy companions orbiting close to the SMBH (i.e., small $a_c$ and large $\mc$). Finally, the last term encompasses the first relativistic corrections, which include the gravitational redshift from both the SMBH and its companion and the transverse Doppler predicted by special relativity (see \cite{Do+19} and \cite{GRAVITY18}).

The dataset used in this analysis is the same as that used in \cite{Do+19} and in \cite{Naoz+20}. It consists of 45 astrometric positional measurements (spanning 24 years) and 115 radial velocities (RVs) spanning 18 years. We used astrometric measurements obtained from the W. M. Keck Observatory by using speckle imaging (a technique to overcome blurring from the atmosphere by taking very short exposures and combining the images with software) from 1995–2005 and adaptive optics (AO) imaging from 2005–2018. These measurements are expressed in the reference frame developed in \cite{sakai:2019ab} and \cite{jia:2019aa} and are publicly available in \cite{Do+19}. In addition, we used RV obtained from six spectroscopic instruments: one from NIRSPEC (Near-Infrared Spectrograph) on Keck, 6 from NIRC2 (Near-Infrared Camera 2) on Keck, 54 from OSIRIS (OH-Suppressing Infra-Red Imaging Spectrograph) on Keck, 9 from NIFS (Near-infrared Integral Field Spectrometer) on Gemini, 4 from IRCS (Infrared Camera and Spectrograph) on Subaru and 41 from SINFONI (SINgle Faint Object Near-IR Investigation) on the Very Large Telescope (VLT). The Keck, Subaru and Gemini data is presented in \cite{Do+19} (see also \cite{Chu+2018}) while the VLT data is reported in \cite{gillessen:2017aa}.

In this analysis, we use Gaussian likelihoods for the RV and the astrometric measurements. The radial velocities are supposed to be independent and normally distributed. We use a Gaussian likelihood for the astrometric measurements, including correlations between the measurements. The covariance matrix for the astrometric measurement depends exponentially on the sky-projected distance between two measurements. It is parameterized by a correlation length $\lambda$ and a mixing parameter $p$, which are both fitted simultaneously with all other model parameters. A detailed discussion of the likelihood used can be found in Sec. 1.5.1 of the Supplementary Materials from \cite{Do+19}. Finally, following the analysis from \cite{Do+19}, we also fit for an offset for the NIRC2 RVs.

  \begin{figure}
  \begin{center} 
    \includegraphics[width=\linewidth]{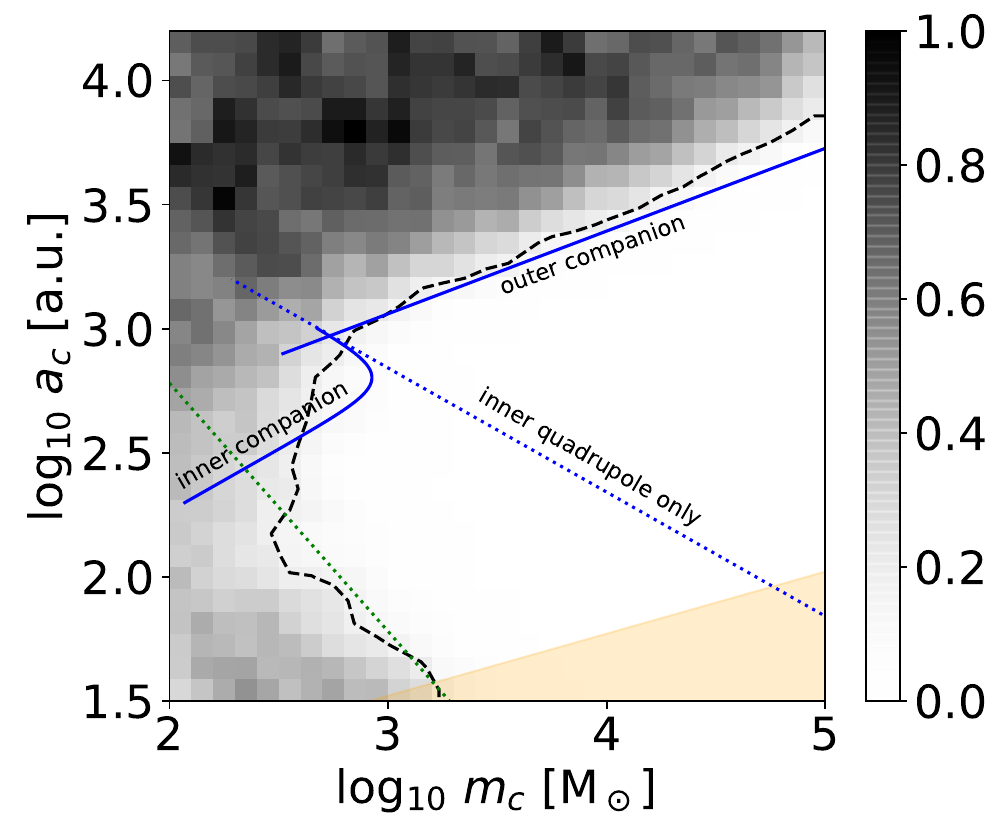}
  \end{center} 
  \caption{  \upshape Bounds on a companion IMBH from numerical simulations.  The grey scale represents the  posterior probability density distribution for the mass of the SMBH companion ($\mc$) and its semi-major axis ($a_c$) marginalized over the other 21 fitted parameters and normalized to 1. The scale of shades at the right of the figure indicates the relative probability. White regions are excluded. The black dashed line corresponds to the 95\% confidence area.The blue curves are the same as in Fig.\ \ref{fig:Bounds}. The dotted green curve corresponds to a $a_c \sim 1/m_c$ behavior.  The tan area is excluded (more precisely is not relevant) as a result of gravitational-wave
damping  for the companion.\label{fig:numerical} }
\end{figure}

In total, in this analysis, we fit simultaneously for 23 parameters: 20 model parameters, one offset for the NIRC2 RVs and 2 parameters to model the correlations between the astrometric measurements. We perform a Bayesian inference for model fitting, using nested sampling to estimate the posterior probability distribution via the MultiNest package \citep{feroz:2008qf,feroz:2009xy}. The resulting 2 dimensional posterior for the parameters $(a_c,\mc)$ marginalized over all the other 21 parameters is presented in Fig.~\ref{fig:numerical}. The overall shape of the confidence area is similar to the one found in \cite{gualandris:2010aa}. We  notice two regimes representing an outer companion and an inner companion, with a turning point corresponding to a semi-major axis of the same order of magnitude as that of S0-2. 
The upper part of the Figure shows an exclusion region (in white) fully compatible with the analytic estimate of Sec.~\ref{sec:outer}, as indicated by the blue curve in Fig.~\ref{fig:numerical}, taken directly from Fig.\ \ref{fig:Bounds}.  It shows the $ac \sim m_c^{1/3}$ behavior expected from quadrupolar perturbations of the S0-2 orbit. 

The lower part of the figure shows quite different behavior.  In addition to excluding high-mass inner companions that would induce quadrupolar perturbations on the orbit of S0-2 (to the right of the line labeled ``inner, quadrupole only''), the results also exclude companions well to the left of that line, corresponding to the region labeled ``b'' in Fig.\ \ref{fig:Bounds}. Here, quadrupolar perturbations of S0-2's orbit are very small, and its orbit serves as a ``fixed reference'' for observing the ``wobble'' of Sgr A* induced by the companion, much as the wobble of stars relative to a fixed background served to discover the first exoplanets. The absence of such an effect in the data serves to exclude companions, for example, with masses as small as $400 \,M_\odot$ at $200$ a.u.  For companions of lower mass, the wobble is too small to be detected, as depicted by solutions with viable companions in the lower left-hand corner of Fig.\ \ref{fig:numerical}.

Other than revealing the potential importance of the wobble effect, the analytic approach does not do a good job of characterizing the bounds in this region of parameter space in detail, largely because it involves integrations over the observation time, which tend to wash out the effect.  By contrast, the numerical integrations incorporate the full time dependence of the wobble and dipole effects, including correlations with other effects. The dotted green line denotes an approximate $a_c \sim m_c^{-1}$ dependence of the bound in this regime, which would suggest the effect of the ``dipole'' term in Eq.\ (\ref{eq:eominnerAst}), but how robust this is remains to be seen.

\section{Other bounds on a companion}
\label{sec:other}

\begin{figure*}
  \begin{center}
    \includegraphics[width=\linewidth]{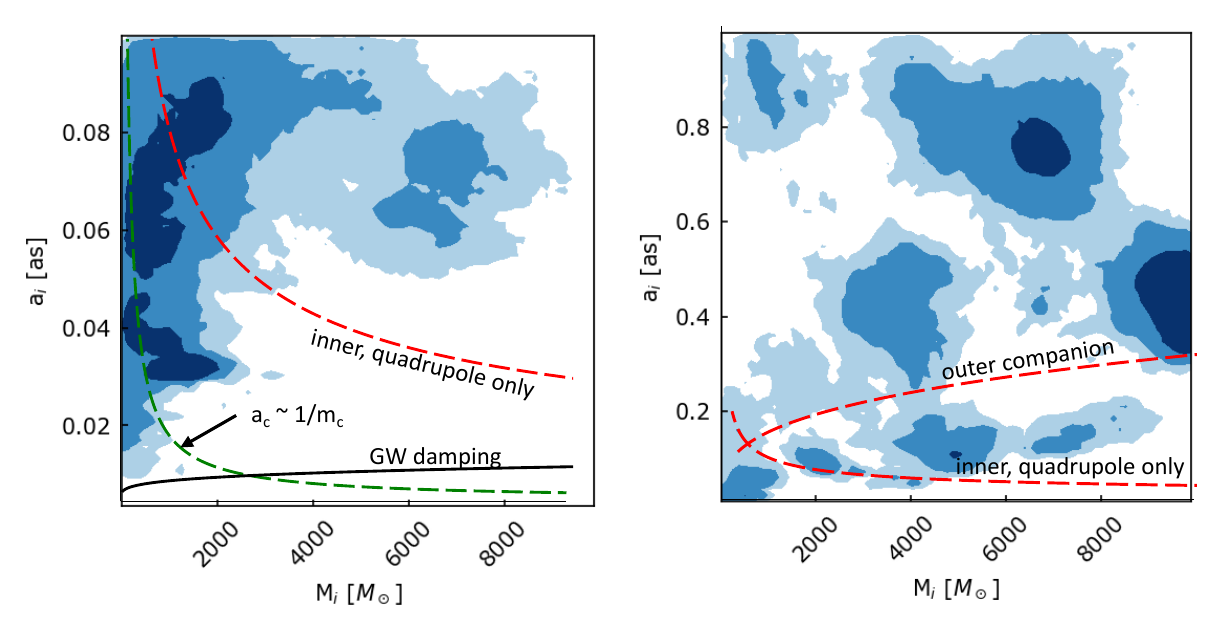}
  \end{center} 
  \caption{  \upshape Comparison between the results of this paper and an analysis by the  \citet{2023A&A...672A..63G}. The left panel corresponds to an inner companion; the right panel to an outer companion.  The shaded blue regions denote companion solutions with 39\%, 86\% and 99\% confidence, from dark to light, respectively. White regions are excluded.  The dashed red curves show our analytic bounds (taken from Fig.\ \ref{fig:Bounds}) for purely quadrupole perturbations.  The dashed green curve (taken from Fig.\ \ref{fig:numerical})  is a rough fit to our numerical results for low-mass inner companions.  Above and to the right of both curves in the left panel, we predict no companions. Between the two dashed red curves in the right panel, we also predict no companions. Background figure reproduced from {\it Astronomy \& Astrophysics} {\bf 672}, A63 (2023) via Creative Commons CC-BY 4.0 license.} \label{fig:GravityCompare} 
\end{figure*}

In a recent paper, the GRAVITY collaboration used numerical simulations to constrain the possibility of a companion IMBH
\citep{2023A&A...672A..63G}. They used four years of astrometric data (2017-21) and 21 years of spectroscopic data, while we used 23 and 18 years' worth, respectively.  Both analyses used the relative separation between S0-2 and Sgr A* as the fundamental variable, and both used similar sets of fitted parameters.  They plotted posterior density distributions for allowed companions, one plot for  an inner companion and one for an outer companion.  For ease of comparison, 
in Fig.\ \ref{fig:GravityCompare} we have reproduced the two main components of their Fig.\ 1  with the left (right) panel corresponding to an inner (outer) companion (note their axes are in linear scale, and $0.125$ arcseconds corresponds to $1020$ a.u.). The shades of blue in the GRAVITY plots correspond to 39\%, 86\% and 99\% confidence (from dark to light).  

In the left panel, we have over-plotted the analytic bound (in red) from pure quadrupole perturbations from an inner companion and the stronger bound (in green) inferred from our numerical simulations, which we have suggested result from the effects of the wobble of Sgr A*.  The gravitational-wave bound is also shown for reference.  In a region where we find no candidate companions -- above and to the right of both the green and red curves -- the GRAVITY analysis seems to find significant numbers of solutions.   

In the right panel of Fig.\ \ref{fig:GravityCompare}, we have plotted in red the analytic bounds for both an outer companion and for an inner companion with quadrupole perturbations.  Between these two curves, we argue that no companions should exist (and our numerical integrations find none), while the GRAVITY analysis shows a number of candidate companions.
Above the ``outer companion'' curve, both analyses agree on the presence of companions compatible with the observations.  The GRAVITY analysis actually goes beyond our study, showing that such outer companions may have a destabilizing effect on the S-star cluster, thus providing additional potential constraints on their existence.

Additional bounds on a hypothetical companion have resulted from limits on the wobble of Sgr A* relative to the distant quasars, and from studies of the effect of a companion on the distribution of inclinations of the S-star cluster.  These primarily exclude high mass companions ($> 2000 \,M_\odot$) exterior to the orbit of S0-2.  Figure 13 of \citet{Gualandris+09} presents a summary of those bounds.
\citet{Zhang+23} placed bounds using a stability criterion for non-hierarchical triple systems, arguing that certain companions could induce changes in the semimajor axis of S0-2 of order unity within the lifetime of the star.  The resulting bound is consistent with the excluded region in Fig.\ \ref{fig:numerical}, though somewhat weaker (i.e. to the right).  \citet{2011ApJ...735...57B} discussed bounds that could be achieved using millimeter very long baseline interferometry.

\section{Conclusions}
\label{sec:conclusions}

We have used astrometric and radial velocity data on the orbit of the star S0-2 to constrain a hypothetical intermediate-mass black hole orbiting the massive black hole Sgr A* at the Galactic center.  We employed a combination of analytic estimates and full numerical integrations of the orbit of S0-2 in the presence of a black-hole binary.  For companions with masses above $10^3 M_\odot$, we found a wedge shaped region in the space of $a_c$ vs. $m_c$ (in log scale) where companions are excluded because their quadrupolar perturbations would induce changes in the orientation of S0-2 larger than the observations allow. Our analytic estimates and numerical simulations were in agreement in this regime.

For lower mass companions inside the orbit of S0-2, analytic estimates suggested that the wobble of Sgr A* about the center of mass of its orbit with the companion would be the main observable effect, but did not give reliable exclusion curves, probably because the averaging methods obscured important short-timescale effects.  However, the numerical simulations verified the importance of the wobble effect and excluded a significant region of the $a_c - m_c$ parameter space, down to masses as small as $400 \, M_\odot$ at $200$ a.u.

\begin{acknowledgments}
 CMW is grateful for the hospitality of the Institut d'Astrophysique de Paris, where parts of this work were carried out, and for partial support from NSF PHY 19-09247 and 22-07681.  SN acknowledges partial support from NASA ATP  80NSSC20K0505, the NSF-AST 2206428 grant, and thanks Howard and Astrid Preston for their generous support. AMG is supported by NSF AST-1909554, the Gordon and Betty Moore Foundation and the Arthur Levine and Lauren Leichtman Chair in Astrophysics at UCLA. AT is supported by NSF PHY-2212983.  We are grateful to Odele Straub for useful comments and for permission to use Fig.\ 1 from \citet{2023A&A...672A..63G}.
\end{acknowledgments}

\bibliographystyle{hapj}
\bibliography{main.bbl}

\begin{thebibliography}{74}
\expandafter\ifx\csname natexlab\endcsname\relax\def\natexlab#1{#1}\fi

\bibitem[{{Bansal} {et~al.}(2017){Bansal}, {Taylor}, {Peck}, {Zavala}, \&
  {Romani}}]{Bansal+17}
{Bansal}, K., {Taylor}, G.~B., {Peck}, A.~B., {Zavala}, R.~T., \& {Romani},
  R.~W. 2017, \apj, 843, 14, 1705.08556

\bibitem[{{Batcheldor} {et~al.}(2010){Batcheldor}, {Robinson}, {Axon},
  {Perlman}, \& {Merritt}}]{Batcheldor+10}
{Batcheldor}, D., {Robinson}, A., {Axon}, D.~J., {Perlman}, E.~S., \&
  {Merritt}, D. 2010, \apjl, 717, L6, 1005.2173

\bibitem[{{Bianchi} {et~al.}(2008){Bianchi}, {Chiaberge}, {Piconcelli},
  {Guainazzi}, \& {Matt}}]{Bianchi+08}
{Bianchi}, S., {Chiaberge}, M., {Piconcelli}, E., {Guainazzi}, M., \& {Matt},
  G. 2008, MNRAS, 386, 105, 0802.0825

\bibitem[{{Bland-Hawthorn} \& {Gerhard}(2016)}]{2016ARA&A..54..529B}
{Bland-Hawthorn}, J., \& {Gerhard}, O. 2016, \araa, 54, 529, 1602.07702

\bibitem[{{Boehle} {et~al.}(2016){Boehle}, {Ghez}, {Sch{\"o}del}, {Meyer},
  {Yelda}, {Albers}, {Martinez}, {Becklin}, {Do}, {Lu}, {Matthews}, {Morris},
  {Sitarski}, \& {Witzel}}]{boehle:2016wu}
{Boehle}, A. {et~al.} 2016, \apj, 830, 17

\bibitem[{{Bogdanovi{\'c}} {et~al.}(2009){Bogdanovi{\'c}}, {Eracleous}, \&
  {Sigurdsson}}]{Bogdanovic+09}
{Bogdanovi{\'c}}, T., {Eracleous}, M., \& {Sigurdsson}, S. 2009, \apj, 697,
  288, 0809.3262

\bibitem[{{Boroson} \& {Lauer}(2009)}]{Boroson+09}
{Boroson}, T.~A., \& {Lauer}, T.~R. 2009, \nat, 458, 53, 0901.3779

\bibitem[{{Broderick} {et~al.}(2011){Broderick}, {Loeb}, \&
  {Reid}}]{2011ApJ...735...57B}
{Broderick}, A.~E., {Loeb}, A., \& {Reid}, M.~J. 2011, \apj, 735, 57, 1104.3146

\bibitem[{{Callegari} {et~al.}(2009){Callegari}, {Mayer}, {Kazantzidis},
  {Colpi}, {Governato}, {Quinn}, \& {Wadsley}}]{Callegari+09}
{Callegari}, S., {Mayer}, L., {Kazantzidis}, S., {Colpi}, M., {Governato}, F.,
  {Quinn}, T., \& {Wadsley}, J. 2009, ApJ-Lett, 696, L89, 0811.0615

\bibitem[{{Capellupo} {et~al.}(2017){Capellupo}, {Haggard}, {Choux},
  {Baganoff}, {Bower}, {Cotton}, {Degenaar}, {Dexter}, {Falcke}, \&
  {Fragile}}]{Capellupo+17}
{Capellupo}, D.~M. {et~al.} 2017, \apj, 845, 35, 1707.01937

\bibitem[{{Chen} \& {Liu}(2013)}]{Chen+13}
{Chen}, X., \& {Liu}, F.~K. 2013, \apj, 762, 95, 1211.4609

\bibitem[{{Chu} {et~al.}(2018){Chu}, {Do}, {Hees}, {Ghez}, {Naoz}, {Witzel},
  {Sakai}, {Chappell}, {Gautam}, {Lu}, \& {Matthews}}]{Chu+2018}
{Chu}, D.~S. {et~al.} 2018, \apj, 854, 12, 1709.04890

\bibitem[{{Comerford} {et~al.}(2009){Comerford}, {Griffith}, {Gerke}, {Cooper},
  {Newman}, {Davis}, \& {Stern}}]{Comerford+09bin}
{Comerford}, J.~M., {Griffith}, R.~L., {Gerke}, B.~F., {Cooper}, M.~C.,
  {Newman}, J.~A., {Davis}, M., \& {Stern}, D. 2009, ApJ-Lett, 702, L82,
  0906.3517

\bibitem[{{Comerford} {et~al.}(2018){Comerford}, {Nevin}, {Stemo},
  {M{\"u}ller-S{\'a}nchez}, {Barrows}, {Cooper}, \& {Newman}}]{Comerford+18}
{Comerford}, J.~M., {Nevin}, R., {Stemo}, A., {M{\"u}ller-S{\'a}nchez}, F.,
  {Barrows}, R.~S., {Cooper}, M.~C., \& {Newman}, J.~A. 2018, \apj, 867, 66,
  1810.11543

\bibitem[{{Deane} {et~al.}(2014){Deane}, {Paragi}, {Jarvis}, {Coriat},
  {Bernardi}, {Fender}, {Frey}, {Heywood}, {Kl{\"o}ckner}, {Grainge}, \&
  {Rumsey}}]{Deane+14}
{Deane}, R.~P. {et~al.} 2014, \nat, 511, 57, 1406.6365

\bibitem[{{Dehnen} \& {Binney}(1998)}]{1998MNRAS.298..387D}
{Dehnen}, W., \& {Binney}, J.~J. 1998, \mnras, 298, 387, astro-ph/9710077

\bibitem[{{Di Matteo} {et~al.}(2005){Di Matteo}, {Springel}, \&
  {Hernquist}}]{DiMatteo+05}
{Di Matteo}, T., {Springel}, V., \& {Hernquist}, L. 2005, \nat, 433, 604,
  astro-ph/0502199

\bibitem[{{Dibi} {et~al.}(2016){Dibi}, {Markoff}, {Belmont}, {Malzac},
  {Neilsen}, \& {Witzel}}]{Dibi+16}
{Dibi}, S., {Markoff}, S., {Belmont}, R., {Malzac}, J., {Neilsen}, J., \&
  {Witzel}, G. 2016, \mnras, 461, 552, 1606.06567

\bibitem[{{Do} {et~al.}(2019){Do}, {Hees}, {Ghez}, {Martinez}, {Chu}, {Jia},
  {Sakai}, {Lu}, {Gautam}, {O{\textquoteright}Neil}, {Becklin}, {Morris},
  {Matthews}, {Nishiyama}, {Campbell}, {Chappell}, {Chen}, {Ciurlo},
  {Dehghanfar}, {Gallego-Cano}, {Kerzendorf}, {Lyke}, {Naoz}, {Saida},
  {Sch{\"o}del}, {Takahashi}, {Takamori}, {Witzel}, \& {Wizinowich}}]{Do+19}
{Do}, T. {et~al.} 2019, Science, 365, 664, 1907.10731

\bibitem[{{Dotti} {et~al.}(2009){Dotti}, {Montuori}, {Decarli}, {Volonteri},
  {Colpi}, \& {Haardt}}]{Dotti+09}
{Dotti}, M., {Montuori}, C., {Decarli}, R., {Volonteri}, M., {Colpi}, M., \&
  {Haardt}, F. 2009, MNRAS, 398, L73, 0809.3446

\bibitem[{{Event Horizon Telescope Collaboration} {et~al.}(2022){Event Horizon
  Telescope Collaboration}, {Akiyama}, {Alberdi}, {Alef}, {Algaba}, {Anantua},
  {Asada}, {Azulay}, {Bach}, {Baczko}, \& et~al.}]{2022ApJ...930L..12E}
{Event Horizon Telescope Collaboration} {et~al.} 2022, \apjl, 930, L12

\bibitem[{{Feroz} \& {Hobson}(2008)}]{feroz:2008qf}
{Feroz}, F., \& {Hobson}, M.~P. 2008, \mnras, 384, 449, 0704.3704

\bibitem[{{Feroz} {et~al.}(2009){Feroz}, {Hobson}, \& {Bridges}}]{feroz:2009xy}
{Feroz}, F., {Hobson}, M.~P., \& {Bridges}, M. 2009, \mnras, 398, 1601,
  0809.3437

\bibitem[{{Fragione} {et~al.}(2020){Fragione}, {Loeb}, {Kremer}, \&
  {Rasio}}]{Fragione+20}
{Fragione}, G., {Loeb}, A., {Kremer}, K., \& {Rasio}, F.~A. 2020, \apj, 897,
  46, 2002.02975

\bibitem[{{Generozov} \& {Madigan}(2020)}]{Generozov+20}
{Generozov}, A., \& {Madigan}, A.-M. 2020, \apj, 896, 137, 2002.10547

\bibitem[{{Ghez} {et~al.}(2003){Ghez}, {Duch{\^e}ne}, {Matthews}, {Hornstein},
  {Tanner}, {Larkin}, {Morris}, {Becklin}, {Salim}, {Kremenek}, {Thompson},
  {Soifer}, {Neugebauer}, \& {McLean}}]{Ghez+03}
{Ghez}, A.~M. {et~al.} 2003, \apjl, 586, L127, astro-ph/0302299

\bibitem[{{Ghez} {et~al.}(2000){Ghez}, {Morris}, {Becklin}, {Tanner}, \&
  {Kremenek}}]{Ghez+00}
{Ghez}, A.~M., {Morris}, M., {Becklin}, E.~E., {Tanner}, A., \& {Kremenek}, T.
  2000, \nat, 407, 349, arXiv:astro-ph/0009339

\bibitem[{{Ghez} {et~al.}(2008){Ghez}, {Salim}, {Weinberg}, {Lu}, {Do}, {Dunn},
  {Matthews}, {Morris}, {Yelda}, {Becklin}, {Kremenek}, {Milosavljevic}, \&
  {Naiman}}]{Ghez+08}
{Ghez}, A.~M. {et~al.} 2008, \apj, 689, 1044, 0808.2870

\bibitem[{{Gillessen} {et~al.}(2009){Gillessen}, {Eisenhauer}, {Trippe},
  {Alexander}, {Genzel}, {Martins}, \& {Ott}}]{Gillessen+09}
{Gillessen}, S., {Eisenhauer}, F., {Trippe}, S., {Alexander}, T., {Genzel}, R.,
  {Martins}, F., \& {Ott}, T. 2009, \apj, 692, 1075, 0810.4674

\bibitem[{{Gillessen} {et~al.}(2017){Gillessen}, {Plewa}, {Eisenhauer}, {Sari},
  {Waisberg}, {Habibi}, {Pfuhl}, {George}, {Dexter}, {von Fellenberg}, {Ott},
  \& {Genzel}}]{gillessen:2017aa}
{Gillessen}, S. {et~al.} 2017, \apj, 837, 30

\bibitem[{{GRAVITY Collaboration} {et~al.}(2020){GRAVITY Collaboration},
  {Abuter}, {Amorim}, {Baub{\"o}ck}, {Berger}, {Bonnet}, {Brand ner},
  {Cardoso}, {Cl{\'e}net}, {de Zeeuw}, {Dexter}, {Eckart}, {Eisenhauer},
  {F{\"o}rster Schreiber}, {Garcia}, {Gao}, {Gendron}, {Genzel}, {Gillessen},
  {Habibi}, {Haubois}, {Henning}, {Hippler}, {Horrobin}, {Jim{\'e}nez-Rosales},
  {Jochum}, {Jocou}, {Kaufer}, {Kervella}, {Lacour}, {Lapeyr{\`e}re}, {Le
  Bouquin}, {L{\'e}na}, {Nowak}, {Ott}, {Paumard}, {Perraut}, {Perrin},
  {Pfuhl}, {Rodr{\'\i}guez-Coira}, {Shangguan}, {Scheithauer}, {Stadler},
  {Straub}, {Straubmeier}, {Sturm}, {Tacconi}, {Vincent}, {von Fellenberg},
  {Waisberg}, {Widmann}, {Wieprecht}, {Wiezorrek}, {Woillez}, {Yazici}, \&
  {Zins}}]{Gravity+20}
{GRAVITY Collaboration} {et~al.} 2020, \aap, 636, L5, 2004.07187

\bibitem[{{GRAVITY Collaboration} {et~al.}(2018){GRAVITY Collaboration},
  {Abuter, R.}, {Amorim, A.}, {Anugu, N.}, {Baub\"ock, M.}, {Benisty, M.},
  {Berger, J. P.}, {Blind, N.}, {Bonnet, H.}, {Brandner, W.}, {Buron, A.},
  {Collin, C.}, {Chapron, F.}, {Cl\'enet, Y.}, {dCoud\'e u Foresto, V.}, {de
  Zeeuw, P. T.}, {Deen, C.}, {Delplancke-Str\"obele, F.}, {Dembet, R.},
  {Dexter, J.}, {Duvert, G.}, {Eckart, A.}, {Eisenhauer, F.}, {Finger, G.},
  {F\"orster Schreiber, N. M.}, {F\'edou, P.}, {Garcia, P.}, {Garcia Lopez,
  R.}, {Gao, F.}, {Gendron, E.}, {Genzel, R.}, {Gillessen, S.}, {Gordo, P.},
  {Habibi, M.}, {Haubois, X.}, {Haug, M.}, {Hau\ss{}mann, F.}, {Henning, Th.},
  {Hippler, S.}, {Horrobin, M.}, {Hubert, Z.}, {Hubin, N.}, {Jimenez Rosales,
  A.}, {Jochum, L.}, {Jocou, L.}, {Kaufer, A.}, {Kellner, S.}, {Kendrew, S.},
  {Kervella, P.}, {Kok, Y.}, {Kulas, M.}, {Lacour, S.}, {Lapeyr\`ere, V.},
  {Lazareff, B.}, {Le Bouquin, J.-B.}, {L\'ena, P.}, {Lippa, M.}, {Lenzen, R.},
  {M\'erand, A.}, {M\"uler, E.}, {Neumann, U.}, {Ott, T.}, {Palanca, L.},
  {Paumard, T.}, {Pasquini, L.}, {Perraut, K.}, {Perrin, G.}, {Pfuhl, O.},
  {Plewa, P. M.}, {Rabien, S.}, {Ram\'{\i}rez, A.}, {Ramos, J.}, {Rau, C.},
  {Rodr\'{\i}guez-Coira, G.}, {Rohloff, R.-R.}, {Rousset, G.},
  {Sanchez-Bermudez, J.}, {Scheithauer, S.}, {Sch\"oller, M.}, {Schuler, N.},
  {Spyromilio, J.}, {Straub, O.}, {Straubmeier, C.}, {Sturm, E.}, {Tacconi, L.
  J.}, {Tristram, K. R. W.}, {Vincent, F.}, {von Fellenberg, S.}, {Wank, I.},
  {Waisberg, I.}, {Widmann, F.}, {Wieprecht, E.}, {Wiest, M.}, {Wiezorrek, E.},
  {Woillez, J.}, {Yazici, S.}, {Ziegler, D.}, \& {Zins, G.}}]{GRAVITY18}
------. 2018, A\&A, 615, L15

\bibitem[{{GRAVITY Collaboration} {et~al.}(2023){GRAVITY Collaboration},
  {Straub}, {Baub{\"o}ck}, {Abuter}, {Aimar}, {Amaro Seoane}, {Amorim},
  {Berger}, {Bonnet}, {Bourdarot}, {Brandner}, {Cardoso}, {Cl{\'e}net},
  {Dallilar}, {Davies}, {de Zeeuw}, {Dexter}, {Drescher}, {Eisenhauer},
  {F{\"o}rster Schreiber}, {Foschi}, {Garcia}, {Gao}, {Gendron}, {Genzel},
  {Gillessen}, {Habibi}, {Haubois}, {Hei{\ss}el}, {Henning}, {Hippler},
  {Horrobin}, {Jochum}, {Jocou}, {Kaufer}, {Kervella}, {Lacour},
  {Lapeyr{\`e}re}, {Le Bouquin}, {L{\'e}na}, {Lutz}, {Ott}, {Paumard},
  {Perraut}, {Perrin}, {Pfuhl}, {Rabien}, {Ribeiro}, {Sadun Bordoni},
  {Scheithauer}, {Shangguan}, {Shimizu}, {Stadler}, {Straubmeier}, {Sturm},
  {Tacconi}, {Vincent}, {von Fellenberg}, {Widmann}, {Wieprecht}, {Wiezorrek},
  {Woillez}, \& {Yazici}}]{2023A&A...672A..63G}
------. 2023, \aap, 672, A63

\bibitem[{{Green} {et~al.}(2010){Green}, {Myers}, {Barkhouse}, {Mulchaey},
  {Bennert}, {Cox}, \& {Aldcroft}}]{Green+10}
{Green}, P.~J., {Myers}, A.~D., {Barkhouse}, W.~A., {Mulchaey}, J.~S.,
  {Bennert}, V.~N., {Cox}, T.~J., \& {Aldcroft}, T.~L. 2010, \apj, 710, 1578,
  1001.1738

\bibitem[{{Gualandris} {et~al.}(2010){Gualandris}, {Gillessen}, \&
  {Merritt}}]{gualandris:2010aa}
{Gualandris}, A., {Gillessen}, S., \& {Merritt}, D. 2010, \mnras, 409, 1146,
  1006.3563

\bibitem[{{Gualandris} \& {Merritt}(2009)}]{Gualandris+09}
{Gualandris}, A., \& {Merritt}, D. 2009, \apj, 705, 361, 0905.4514

\bibitem[{{G{\"u}rkan} \& {Rasio}(2005)}]{Grkan+20}
{G{\"u}rkan}, M.~A., \& {Rasio}, F.~A. 2005, \apj, 628, 236, astro-ph/0412452

\bibitem[{{Habibi} {et~al.}(2017){Habibi}, {Gillessen}, {Martins},
  {Eisenhauer}, {Plewa}, {Pfuhl}, {George}, {Dexter}, {Waisberg}, {Ott}, {von
  Fellenberg}, {Baub{\"o}ck}, {Jimenez-Rosales}, \& {Genzel}}]{habibi:2017aa}
{Habibi}, M. {et~al.} 2017, \apj, 847, 120, 1708.06353

\bibitem[{{Hansen} \& {Milosavljevi{\'c}}(2003)}]{Hansen+03}
{Hansen}, B.~M.~S., \& {Milosavljevi{\'c}}, M. 2003, ApJ-Lett, 593, L77,
  arXiv:astro-ph/0306074

\bibitem[{{Hees} {et~al.}(2017){Hees}, {Do}, {Ghez}, {Martinez}, {Naoz},
  {Becklin}, {Boehle}, {Chappell}, {Chu}, {Dehghanfar}, {Kosmo}, {Lu},
  {Matthews}, {Morris}, {Sakai}, {Sch{\"o}del}, \& {Witzel}}]{hees:2017aa}
{Hees}, A. {et~al.} 2017, Physical Review Letters, 118, 211101

\bibitem[{{Hopkins} {et~al.}(2006){Hopkins}, {Hernquist}, {Cox}, {Di Matteo},
  {Robertson}, \& {Springel}}]{Hopkins+06}
{Hopkins}, P.~F., {Hernquist}, L., {Cox}, T.~J., {Di Matteo}, T., {Robertson},
  B., \& {Springel}, V. 2006, ApJS, 163, 1, astro-ph/0506398

\bibitem[{{Jia} {et~al.}(2019){Jia}, {Lu}, {Sakai}, {Gautam}, {Do}, {Hosek},
  {Service}, {Ghez}, {Gallego-Cano}, {Sch{\"o}del}, {Hees}, {Morris},
  {Becklin}, \& {Matthews}}]{jia:2019aa}
{Jia}, S. {et~al.} 2019, \apj, 873, 9

\bibitem[{{Kharb} {et~al.}(2017){Kharb}, {Lal}, \& {Merritt}}]{Kharb+17}
{Kharb}, P., {Lal}, D.~V., \& {Merritt}, D. 2017, Nature Astronomy, 1, 727,
  1709.06258

\bibitem[{{Komossa} {et~al.}(2003){Komossa}, {Burwitz}, {Hasinger}, {Predehl},
  {Kaastra}, \& {Ikebe}}]{Komossa+03}
{Komossa}, S., {Burwitz}, V., {Hasinger}, G., {Predehl}, P., {Kaastra}, J.~S.,
  \& {Ikebe}, Y. 2003, ApJ-Lett, 582, L15, astro-ph/0212099

\bibitem[{{Komossa} {et~al.}(2008){Komossa}, {Zhou}, \& {Lu}}]{Komossa+08}
{Komossa}, S., {Zhou}, H., \& {Lu}, H. 2008, ApJ-Lett, 678, L81, 0804.4585

\bibitem[{{Kozai}(1962)}]{Kozai}
{Kozai}, Y. 1962, \aj, 67, 591

\bibitem[{{Li} {et~al.}(2016){Li}, {Wang}, {Ho}, {Lu}, {Qiu}, {Du}, {Hu},
  {Huang}, {Zhang}, {Wang}, \& {Bai}}]{Li+16}
{Li}, Y.-R. {et~al.} 2016, \apj, 822, 4, 1602.05005

\bibitem[{{Lidov}(1962)}]{Lidov}
{Lidov}, M.~L. 1962, planss, 9, 719

\bibitem[{Liu {et~al.}(2016)Liu, Eracleous, \& Halpern}]{Liu+17}
Liu, J., Eracleous, M., \& Halpern, J.~P. 2016, The Astrophysical Journal, 817,
  42

\bibitem[{{Liu} {et~al.}(2010){Liu}, {Greene}, {Shen}, \&
  {Strauss}}]{Liu+10kpc}
{Liu}, X., {Greene}, J.~E., {Shen}, Y., \& {Strauss}, M.~A. 2010, ApJ-Lett,
  715, L30, 1003.3467

\bibitem[{{Liu} {et~al.}(2014){Liu}, {Shen}, {Bian}, {Loeb}, \&
  {Tremaine}}]{Liu+14}
{Liu}, X., {Shen}, Y., {Bian}, F., {Loeb}, A., \& {Tremaine}, S. 2014, \apj,
  789, 140, 1312.6694

\bibitem[{{Lu} {et~al.}(2013){Lu}, {Do}, {Ghez}, {Morris}, {Yelda}, \&
  {Matthews}}]{Lu+13}
{Lu}, J.~R., {Do}, T., {Ghez}, A.~M., {Morris}, M.~R., {Yelda}, S., \&
  {Matthews}, K. 2013, \apj, 764, 155, 1301.0540

\bibitem[{{Maillard} {et~al.}(2004){Maillard}, {Paumard}, {Stolovy}, \&
  {Rigaut}}]{Maillard+04}
{Maillard}, J.~P., {Paumard}, T., {Stolovy}, S.~R., \& {Rigaut}, F. 2004, \aap,
  423, 155, arXiv:astro-ph/0404450

\bibitem[{{Naoz}(2016)}]{Naoz16}
{Naoz}, S. 2016, \araa, 54, 441, 1601.07175

\bibitem[{{Naoz} {et~al.}(2017){Naoz}, {Li}, {Zanardi}, {de El{\'\i}a}, \& {Di
  Sisto}}]{Naoz+17}
{Naoz}, S., {Li}, G., {Zanardi}, M., {de El{\'\i}a}, G.~C., \& {Di Sisto},
  R.~P. 2017, \aj, 154, 18, 1701.03795

\bibitem[{{Naoz} {et~al.}(2020){Naoz}, {Will}, {Ramirez-Ruiz}, {Hees}, {Ghez},
  \& {Do}}]{Naoz+20}
{Naoz}, S., {Will}, C.~M., {Ramirez-Ruiz}, E., {Hees}, A., {Ghez}, A.~M., \&
  {Do}, T. 2020, \apjl, 888, L8, 1912.04910

\bibitem[{{Pesce} {et~al.}(2018){Pesce}, {Braatz}, {Condon}, \&
  {Greene}}]{Pesce+18}
{Pesce}, D.~W., {Braatz}, J.~A., {Condon}, J.~J., \& {Greene}, J.~E. 2018,
  \apj, 863, 149, 1807.04598

\bibitem[{Poisson \& Will(2014)}]{PW2014}
Poisson, E., \& Will, C.~M. 2014, Gravity: Newtonian, Post-Newtonian,
  Relativistic (Cambridge: Cambridge University Press)

\bibitem[{{Rashkov} \& {Madau}(2013)}]{RM13}
{Rashkov}, V., \& {Madau}, P. 2013, ArXiv e-prints, 1303.3929

\bibitem[{{Robertson} {et~al.}(2006){Robertson}, {Bullock}, {Cox}, {Di Matteo},
  {Hernquist}, {Springel}, \& {Yoshida}}]{Robertson+06}
{Robertson}, B., {Bullock}, J.~S., {Cox}, T.~J., {Di Matteo}, T., {Hernquist},
  L., {Springel}, V., \& {Yoshida}, N. 2006, \apj, 645, 986, astro-ph/0503369

\bibitem[{{Rodriguez} {et~al.}(2006){Rodriguez}, {Taylor}, {Zavala}, {Peck},
  {Pollack}, \& {Romani}}]{Rodriguez+06}
{Rodriguez}, C., {Taylor}, G.~B., {Zavala}, R.~T., {Peck}, A.~B., {Pollack},
  L.~K., \& {Romani}, R.~W. 2006, \apj, 646, 49, astro-ph/0604042

\bibitem[{{Rose} {et~al.}(2022){Rose}, {Naoz}, {Sari}, \& {Linial}}]{Rose+22}
{Rose}, S.~C., {Naoz}, S., {Sari}, R., \& {Linial}, I. 2022, \apjl, 929, L22,
  2201.00022

\bibitem[{{Rose} {et~al.}(2023){Rose}, {Naoz}, {Sari}, \& {Linial}}]{Rose+23}
------. 2023, arXiv e-prints, arXiv:2304.10569, 2304.10569

\bibitem[{{Runnoe} {et~al.}(2017){Runnoe}, {Eracleous}, {Pennell}, {Mathes},
  {Boroson}, {Sigursson}, {Bogdanovc}, {Halpern}, {Liu}, \&
  {Brown}}]{Runnoe+17}
{Runnoe}, J.~C. {et~al.} 2017, MNRAS, 468, 1683, 1702.05465

\bibitem[{{Sakai} {et~al.}(2019){Sakai}, {Lu}, {Ghez}, {Jia}, {Do}, {Witzel},
  {Gautam}, {Hees}, {Becklin}, {Matthews}, \& {Hosek}}]{sakai:2019ab}
{Sakai}, S. {et~al.} 2019, \apj, 873, 65

\bibitem[{{Sillanpaa} {et~al.}(1988){Sillanpaa}, {Haarala}, {Valtonen},
  {Sundelius}, \& {Byrd}}]{Sillanpaa+88}
{Sillanpaa}, A., {Haarala}, S., {Valtonen}, M.~J., {Sundelius}, B., \& {Byrd},
  G.~G. 1988, \apj, 325, 628

\bibitem[{{Smith} {et~al.}(2010){Smith}, {Shields}, {Bonning}, {McMullen},
  {Rosario}, \& {Salviander}}]{Smith+10}
{Smith}, K.~L., {Shields}, G.~A., {Bonning}, E.~W., {McMullen}, C.~C.,
  {Rosario}, D.~J., \& {Salviander}, S. 2010, \apj, 716, 866, 0908.1998

\bibitem[{{Stemo} {et~al.}(2020){Stemo}, {Comerford}, {Barrows}, {Stern},
  {Assef}, {Griffith}, \& {Schechter}}]{Stemo+20}
{Stemo}, A., {Comerford}, J.~M., {Barrows}, R.~S., {Stern}, D., {Assef}, R.~J.,
  {Griffith}, R.~L., \& {Schechter}, A. 2020, arXiv e-prints, arXiv:2011.10051,
  2011.10051

\bibitem[{{The LIGO Scientific Collaboration} {et~al.}(2020{\natexlab{a}}){The
  LIGO Scientific Collaboration}, {the Virgo Collaboration}, {Abbott},
  {Abbott}, {Abraham}, {Acernese}, {Ackley}, {Adams}, {Adhikari}, {Adya},
  {Affeldt}, {Agathos}, {Agatsuma}, {Aggarwal}, {Aguiar}, {Aich}, {Aiello},
  {Ain}, {Zucker}, \& {Zweizig}}]{GW190521a+20}
{The LIGO Scientific Collaboration} {et~al.} 2020{\natexlab{a}}, arXiv
  e-prints, arXiv:2009.01075, 2009.01075

\bibitem[{{The LIGO Scientific Collaboration} {et~al.}(2020{\natexlab{b}}){The
  LIGO Scientific Collaboration}, {the Virgo Collaboration}, {Abbott},
  {Abbott}, {Abraham}, {Acernese}, {Ackley}, {Adams}, {Adhikari}, {Adya},
  {Affeldt}, {Agathos}, {Agatsuma}, {Aggarwal}, {Aguiar}, {Aich}, {Aiello},
  {Ain}, {Ajith}, {Akcay}, {Allen}, {Allocca}, {Altin}, {Amato}, {Anand},
  {Ananyeva}, {Anderson}, {Anderson}, {Angelova}, {Ansoldi}, {Antier},
  {Appert}, {Arai}, {Araya}, {Areeda}, {Ar{\`e}ne}, {Arnaud}, {Aronson},
  {Arun}, {Asali}, {Ascenzi}, {Ashton}, {Aston}, {Astone}, {Aubin}, {Aufmuth},
  {AultONeal}, {Austin}, {Avendano}, {Babak}, {Bacon}, {Badaracco}, {Bader},
  {Bae}, {Baer}, {Baird}, {Baldaccini}, {Ballardin}, {Ballmer}, {Zucker}, \&
  {Zweizig}}]{GW190521b+20}
------. 2020{\natexlab{b}}, arXiv e-prints, arXiv:2009.01190, 2009.01190

\bibitem[{{Witzel} {et~al.}(2018){Witzel}, {Martinez}, {Hora}, {Willner},
  {Morris}, {Gammie}, {Becklin}, {Ashby}, {Baganoff}, \& {Carey}}]{Witzel+18}
{Witzel}, G. {et~al.} 2018, \apj, 863, 15, 1806.00479

\bibitem[{{Zanardi} {et~al.}(2017){Zanardi}, {de El{\'\i}a}, {Di Sisto},
  {Naoz}, {Li}, {Guilera}, \& {Brunini}}]{Zanardi+17}
{Zanardi}, M., {de El{\'\i}a}, G.~C., {Di Sisto}, R.~P., {Naoz}, S., {Li}, G.,
  {Guilera}, O.~M., \& {Brunini}, A. 2017, \aap, 605, A64, 1701.03865

\bibitem[{{Zhang} {et~al.}(2023){Zhang}, {Naoz}, \& {Will}}]{Zhang+23}
{Zhang}, E., {Naoz}, S., \& {Will}, C.~M. 2023, arXiv e-prints,
  arXiv:2301.08271, 2301.08271

\bibitem[{{Zheng} {et~al.}(2020){Zheng}, {Lin}, \& {Mao}}]{Zheng+20}
{Zheng}, X., {Lin}, D. N.~C., \& {Mao}, S. 2020, arXiv e-prints,
  arXiv:2011.04653, 2011.04653

\end{thebibliography}
 
\end{document}